\documentclass[manuscript]{aastex}

\slugcomment{Draft current of \today}
\usepackage{amssymb, amsmath}

\shorttitle{NEW OB STARS IN CARINA}
\shortauthors{Alexander et al.}

\begin{document}

\title{A CATALOG OF NEW SPECTROSCOPICALLY CONFIRMED MASSIVE OB STARS IN CARINA}

\author{Michael J.\ Alexander\altaffilmark{1}, Richard J.\ Hanes} 
\affil{Department of Physics, Lehigh University, 16 Memorial Drive East, Bethlehem, PA 18015, USA}
\email{alexamic@lafayette.edu, rjh314@lehigh.edu}

\author{Matthew S.\ Povich}
\affil{Department of Physics \& Astronomy, California State Polytechnic University Polytechnic, Pomona, CA 91768, USA}
\email{mspovich@cpp.edu}

\and

\author{M.\ Virginia McSwain}
\affil{Department of Physics, Lehigh University, 16 Memorial Drive East, Bethlehem, PA 18015, USA}
\email{mcswain@lehigh.edu}

\altaffiltext{1}{Current address: Department of Physics, Lafayette College, 730 High Street, Easton, PA 18042, USA}

\begin{abstract}

The Carina star-forming region is one of the largest in the Galaxy, and its massive star population is still being unveiled. The large number of stars combined with high, and highly variable, interstellar extinction makes it inherently difficult to find OB stars in this type of young region. 
We present the results of a spectroscopic campaign to study the massive star population of the Carina Nebula, with the primary goal to confirm or reject previously identified Carina OB star candidates. A total of 141 known O- and B-type stars and 94 candidates were observed, of which 73 candidates had a high enough signal-to-noise ratio to classify. We find 23 new OB stars within the Carina Nebula, a 32\%\ confirmation rate. One of the new OB stars has blended spectra and is suspected to be a double-lined spectroscopic binary (SB2).  We also reclassify the spectral types of the known OB stars and discover nine new SB2s among this population.  Finally, we discuss the spatial distribution of these new OB stars relative to known structures in the Carina NSebula.

\end{abstract}


\section{Introduction}

The Carina Nebula is one of the most active star-forming regions in the Milky Way. It is host to more than 200 massive OB stars \citep{gfs11} and over 1400 young stellar objects (YSOs; \citealt{psm11}). Many of the O stars are contained within several dense clusters, including Trumpler (Tr) 14, Tr 15, Tr 16, Bochum (Bo) 10, and Bo 11. Because the nebula is so large (more than 1~deg$^{2}$ on the sky) most spectroscopic studies have focused on the clusters \citep{mdw01}, and not the field, so the census of massive stars in the region is likely incomplete, especially for less luminous late O and early B spectral types.

\citet{ptb11}, hereafter P11, employed data from the \textit{Chandra} Carina Complex Project (CCCP; \citealt{tbc11}) to propose new candidate massive stars in the region. Massive stars may emit X-rays through different processes, notably embedded wind shocks and colliding wind shocks in binary systems. \citet{gfs11} found that 68/70 O stars and 61/127 B0--B3 stars have detections in the CCCP catalog, so P11 restricted their search to 3444 sources that have X-ray detections and no IR excess (to eliminate pre-main sequence [PMS] stars). In total, P11 identified 94 new candidate massive OB stars within the CCCP survey area (1.42~deg$^{2}$). \citet{spw10} also compiled a list of `extended red objects' (EROs) that have extremely red colors in \textit{Spitzer} images. Many EROs are associated with known massive OB stars, and a few display a prominent, resolved bow-shock morphology. An extension of this work by \citet{sps15} revealed additional EROs as well as confirming new massive stars through spectroscopic observations.  

In this paper, we present results from a campaign to obtain spectroscopic classifications for the P11 candidate O star sample and several of the EROs. Section~\ref{obs-sec} describes the observations and data reduction, Section~\ref{spec-sec} details the spectral classification, and Section~\ref{disc-sec} presents a discussion of notable results.

\section{Observations}\label{obs-sec}

We obtained optical spectroscopic observations of 141 known and 94 candidate OB stars and EROs over three nights and two observing runs at the 4~m Anglo-Australian Telescope (AAT). The first night of observations was taken on 2013 March 18, and the second run was on 2014 April 07-08. Unfortunately, the observations from 2013 were cut short owing to bad weather and wildfires on Siding Spring Mountain, where the telescope is located.

\begin{deluxetable}{rrrrrr}
\tabletypesize{\scriptsize}
\tablecaption{AAT observations  \label{table-obs}}
\tablewidth{0pt}
\tablehead{\colhead{Date} & \colhead{Time} & \colhead{Targets} & \colhead{Airmass} & \colhead{Exp.} & \colhead{$\lambda_{\mathrm{cen}}$} \\
           \multicolumn{2}{c}{(UT)} & \colhead{ } & \colhead{} & \colhead{(s)} & \colhead{(\AA)}
}

\startdata
2013-03-18 & 02:02 & bright & 1.62 & 300 & 4100 \\
2013-03-18 & 02:09 & bright & 1.59 & 300 & 4100 \\
2013-03-18 & 02:16 & bright & 1.56 & 300 & 4100 \\
2013-03-18 & 02:47 & faint & 1.41 & 1800 & 4100 \\
2013-03-18 & 03:19 & faint & 1.32 & 1800 & 4100 \\
\hline \\
2014-04-07 & 13:05 & bright & 1.17 &  300 & 4375 \\
2014-04-07 & 13:15 & bright & 1.18 & 1200 & 4375 \\
2014-04-07 & 13:37 & bright & 1.21 & 1200 & 4375 \\
2014-04-07 & 13:59 & bright & 1.24 & 1200 & 4375 \\
2014-04-07 & 14:21 & bright & 1.28 & 1200 & 4375 \\
2014-04-07 & 15:09 &  faint & 1.39 & 1800 & 4410 \\
2014-04-07 & 15:41 &  faint & 1.49 & 1800 & 4410 \\
\hline \\
2014-04-08 &  8:54 &  faint & 1.30 & 1800 & 4410 \\
2014-04-08 &  9:26 &  faint & 1.24 & 1800 & 4410 \\
2014-04-08 &  9:58 &  faint & 1.22 & 1800 & 4410 \\
2014-04-08 & 10:40 & bright & 1.16 & 1200 & 4410 \\
2014-04-08 & 11:02 & bright & 1.15 & 1200 & 4410 \\
2014-04-08 & 11:27 & bright & 1.14 & 1200 & 4410 \\
2014-04-08 & 12:02 &  faint & 1.14 & 1800 & 4410 \\
2014-04-08 & 12:34 &  faint & 1.15 &  948 & 4410 \\
2014-04-08 & 12:58 &  faint & 1.17 & 1800 & 4410 \\
2014-04-08 & 13:30 &  faint & 1.20 & 1800 & 4410 \\
2014-04-08 & 14:02 &  faint & 1.25 & 1800 & 4410 \\
\enddata
\end{deluxetable}

Table~\ref{table-obs} lists the telescope/instrument setup for individual exposures. All data were taken using the AAOmega two-degree field (2dF) fiber-fed multi-object spectrograph with the medium-resolution ($R\sim8900$) 3200B grating. Because of the wide range of apparent magnitudes, target stars were split into separate `bright' and `faint' groups for observation to maximize the signal-to-noise (S/N) and prevent saturation. A total of 392 fibers can be positioned, and the wavelength coverage varies slightly owing to slight differences in dispersion across the chip. The spectrograph was also set up with two different central wavelengths (4100 and 4410 \AA) in order to maximize the number of lines for spectral typing.  A setting with central wavelength of 4375 \AA\ was tried for some exposures, but we found that this setting omitted the \ion{He}{1} $\lambda$4471 line for many targets due to the variable dispersion across the chip. For each setup, the goal was to achieve a target S/N $\ga 100$ for most of the observed stars.

The data were reduced using the \textsc{dohydra} package within IRAF along with a custom sky subtraction program that accounted for the variable dispersion to remove telluric features and partially remove nebular emission. All spectra were wavelength calibrated using a set of arcs including CuAr, FeAr, and NeAr lamps.

\section{Spectroscopic Classification}\label{spec-sec}

We used the spectral atlas of \citet{wf90} to identify massive stars within the AAT sample. The classifications of O stars relied on the relative strengths of \ion{He}{2} $\lambda$4200 and \ion{He}{1}/II $\lambda$4026 in the short-wavelength spectra. B stars are classified primarily using the strength of He I lines (max at B2) and strength of \ion{C}{3} + \ion{O}{2} blend near $\lambda$4070 and \ion{Si}{4} $\lambda$4089.  Luminosity class I and III were assigned primarily by the strength of \ion{Si}{4} $\lambda$4089 and Balmer line absorption. For the longer-wavelength spectra, the primary identifiers were the strengths of the \ion{He}{1}~$\lambda$$\lambda$4387 and 4471 lines, the ratio of \ion{He}{1} $\lambda$4471 to \ion{Mg}{2} $\lambda$4481 (when present), and H$\gamma$. Later-type stars were identified using Gray's Digital Spectral Classification Atlas v1.02.\footnote{http://ned.ipac.caltech.edu/level5/Gray/Gray\_contents.html} These stars are easy to separate from OB types as they are rich in lines from metals and show prominent absorption features like the CH G band.

Table~\ref{table-obk} lists 141 previously known OB stars associated with the Carina Nebula. Column 1 gives the stars' common name, columns 2 and 3 are the R.A. and decl. (J2000), column 4 is the spectral type estimated from the 4100 \AA\ central wavelength observations, column 5 is the spectral type estimated from the 4375/4410 \AA\ central wavelength observations, column 6 is the spectral type listed in the SIMBAD database,\footnote{http://simbad.u-strasbg.fr/simbad/} and column 7 provides additional notes on interesting sources. The spectral types we obtain generally agree to better than two subtypes of the literature values.  Larger differences in spectral type may be due to previously unknown binarity or other variability.  We also note that two Be stars identified in the literature (LS 1914 and HD 93683) did not show any evidence of Balmer or other line emission in our spectra.  Three of our target OB stars (Tr 16-122, Tr 16-124, and Tr 16-127) are misidentified as belonging to the Tr 14 cluster by \citet{gfs11}.  

From the sample of 94 candidate OB stars from P11, 92 of them were included as targets in the AAT observations; OBc~51~and~86 were not observed. Two EROs with unknown spectral types were also included as OB candidates in our observations.  Of these 94 candidate OB stars, 21 stars had spectra whose S/N was too low to make a meaningful judgment on spectral types. We spectroscopically confirm 23 of the candidate stars, which are listed in Table~\ref{table-obcob}. Column 1 lists the OB candidate number from P11 or ERO number from \citet{sps15} , columns 2 and 3 are coordinates, and columns 4, 5, and 6 are the spectral types for the two wavelength ranges and notes as in Table~\ref{table-obk}. Table~\ref{table-obcfgk} has a similar format to Table~\ref{table-obcob} and lists the 71 spectroscopically rejected or undetected candidate OB stars. 

\begin{figure}
\vspace{-.5in}
\epsscale{1}
\plotone{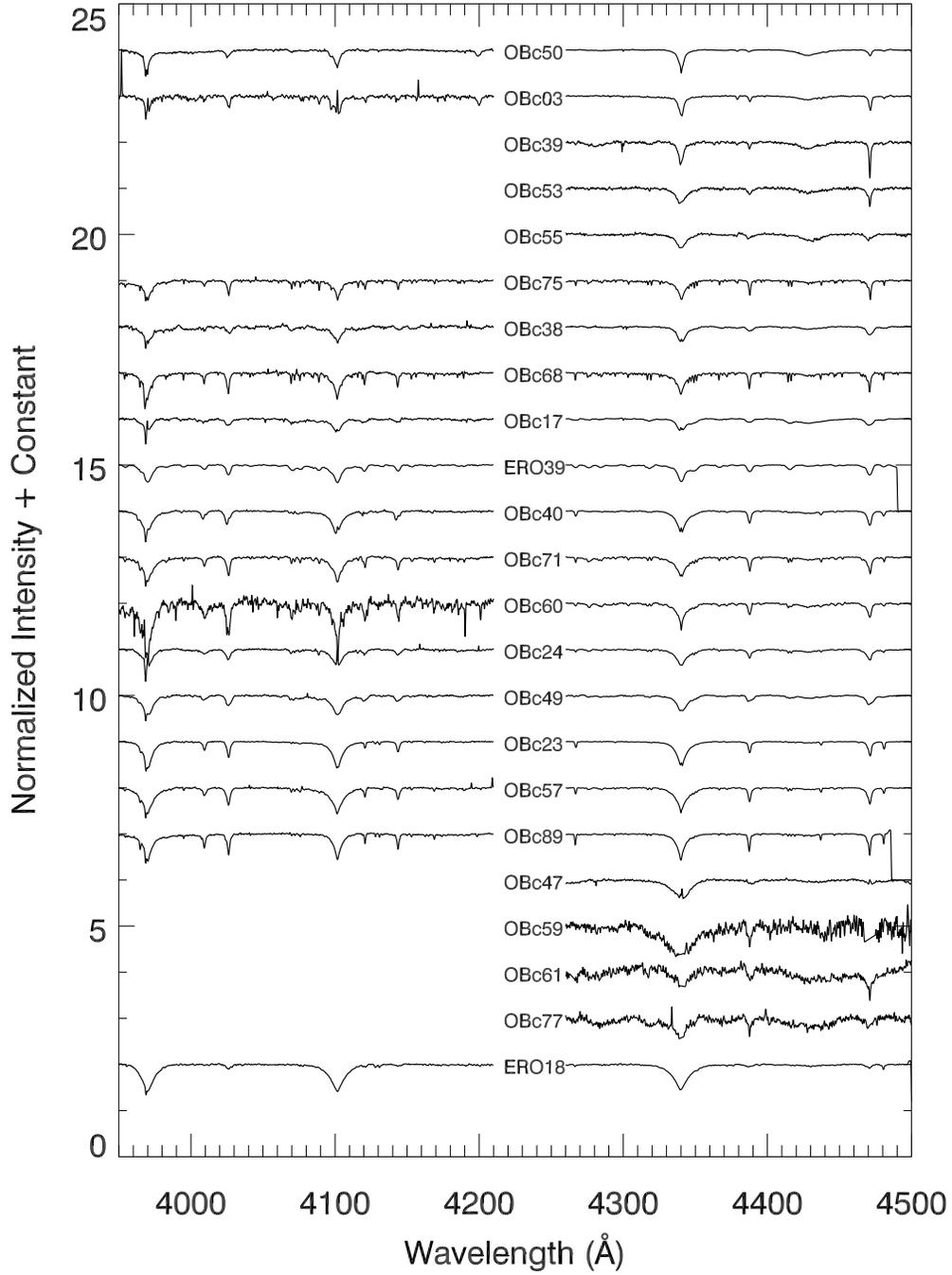}
\caption{Spectra of confirmed OB candidates for both observed wavelength ranges. Stars are shown in approximate order of earlier to later spectral types (top to bottom).  The spectra have been smoothed and cleaned for glitches for display purposes.  \label{fig-obcplot}}
\end{figure}

Figure~\ref{fig-obcplot} plots spectra for the 23 confirmed OB candidates, shown in approximate order of decreasing spectral type from top to bottom. Each stars' identifier is listed alongside its spectrum. A few of the shorter wavelength spectra are missing as these did not have sufficient S/N or were too faint to be detected during the 2013 run. 

\section{Discussion}\label{disc-sec}

\begin{figure}[htp]
\epsscale{.9}
\plotone{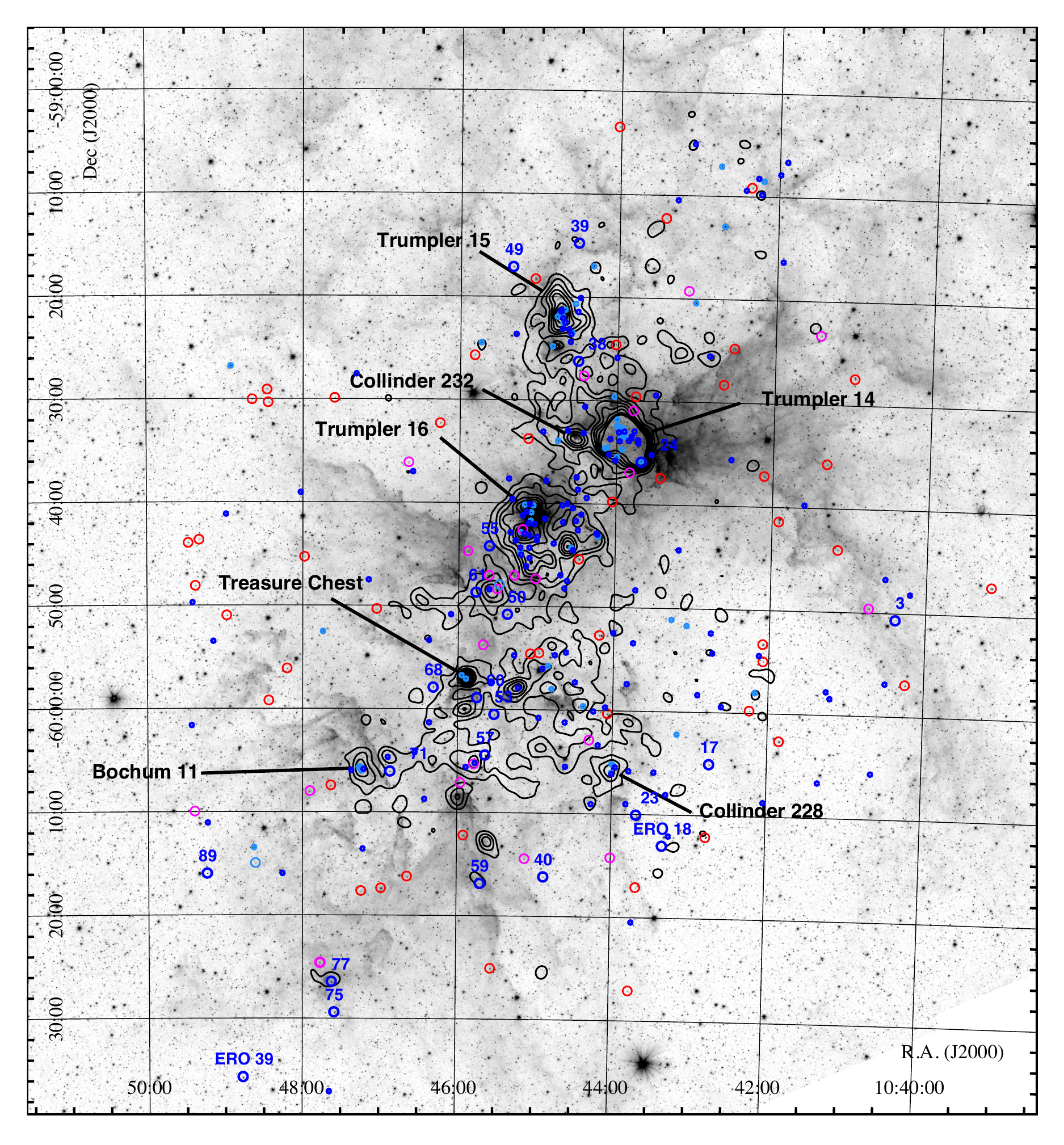}
\caption{\textit{Spitzer} 3.6 \micron\ image of the Carina star-forming region. Larger and smaller circles mark the locations of candidate OB stars and previously classified OB stars, respectively. OB stars spectroscopically classified/confirmed by this work are colored dark blue, rejected OB candidates are red, nondetections are magenta, and unobserved are light blue. Black contours show the stellar density traced by the X-ray point-source population \citep{feigelson2011}; several well-known clusters are annotated and visible as local peaks in the contour map. \label{fig-spatial}}
\end{figure}

We have observed nearly all of the candidate OB stars from P11 and confirm that 21/92 are bona fide OB stars with a spectral range of O6 to B2.  All of the EROs were confirmed, for a total of 23 new OB stars in Carina.  A majority of the observed stars are G/K types, with a few A and F stars. Simulations of X-ray-emitting sources within the CCCP area estimate about 5000 contaminating sources not associated with the Carina star-forming region: 2500 active galactic nuclei, 1800 foreground stars, and 900 background stars \citep{gbf11}. Of these, background K and M giants are the brightest in the IR and are most likely to be misidentified as reddened OB stars. Between the estimated number of X-ray contaminants and the IR detection limits, P11 estimated the OB confirmation rate to be $\gtrsim60\%$; however, the observed rate is only 32\%.  Figure~\ref{fig-spatial} shows the distribution of spectroscopically confirmed OB candidate stars, as well as the the locations of known star clusters.

We do not believe that the rejected OB candidates could be intermediate-mass PMS stars within the Carina complex.  The minimum luminosity cutoff for an OB candidate corresponds to a main-sequence spectral type of B1 V. This was chosen by P11 to minimize contamination of the OB candidate sample from the bright tail of the PMS stellar population. As P11 showed (see their Figure 2), the locus of SED model fits to the OB candidates overlaps the theoretical PMS tracks only for extremely young ages (${<}0.5$~Myr), when low- or intermediate-mass stars are expected to show strong IR excess emission from circumstellar dust disks or protostellar envelopes. 

\begin{figure}[htp]
\epsscale{0.75}
\plotone{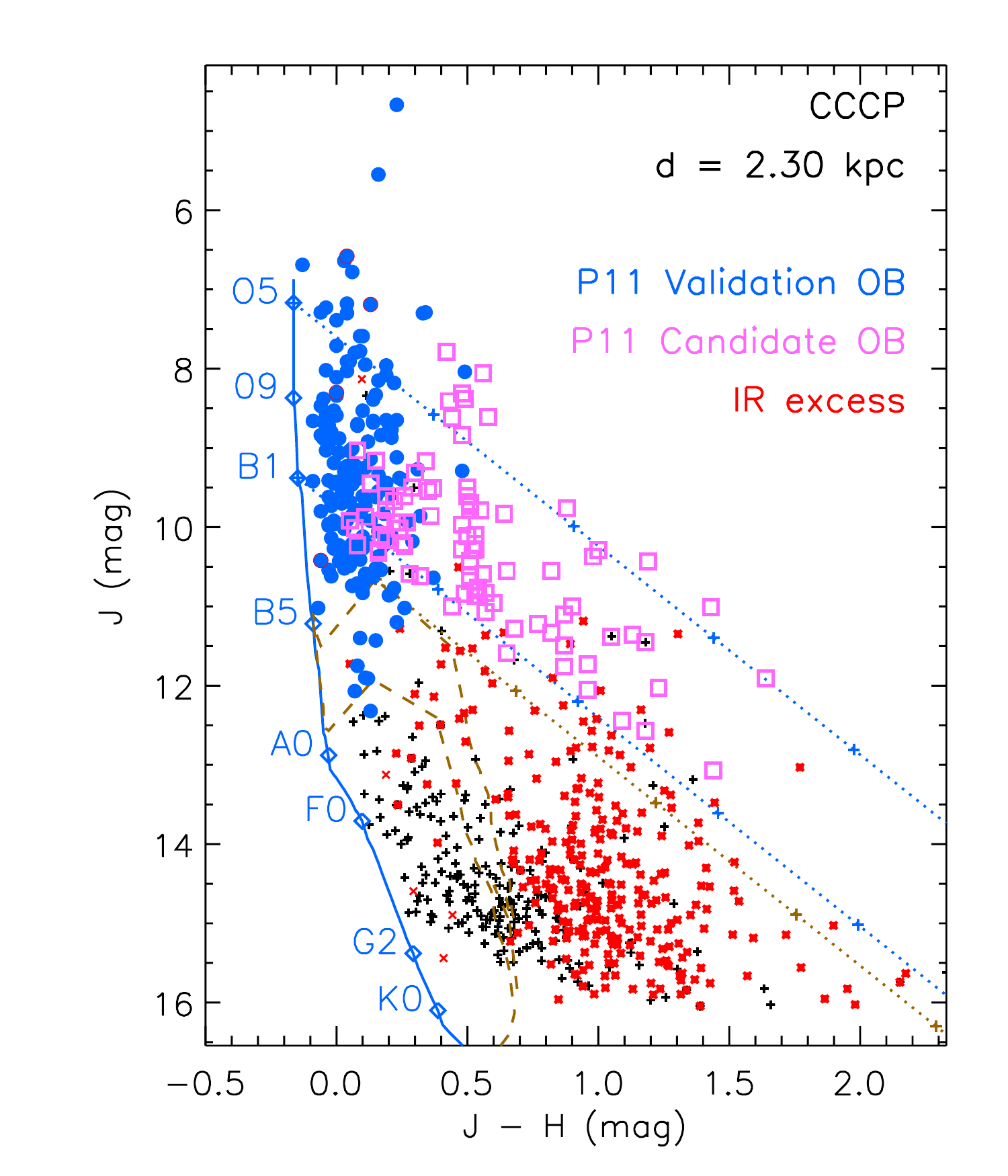}
\caption{$J$ vs $J-H$ near-IR color-magnitude diagram showing Two Micron All Sky Survey sources matched to X-ray-detected PMS and OB stars in the CCCP catalogs. This plot includes the ``validation sample'' of known OB stars (blue filled circles) and candidate OB stars (lavender squares) from P11, plus the near-IR bright ($J < 16$~mag) subset of the CCCP MPCM population, which includes both YSOs with IR excess emission from disks and envelopes (red symbols) and ``diskless'' PMS stars (black plus signs). The main-sequence is plotted as a blue curve with spectral subclasses labeled, with 1 and 3~Myr PMS isochrones \citep{siess2000} plotted as dashed brown curves. Three reddening vectors are marked by plus signs at invervals corresponding to $A_V= 5$~mag.  The PMS stars are not a significant contaminant of the candidate OB stars. 
\label{NIRCMD} }
\end{figure}

We show the relative locations of known and candidate OB stars and the near-IR bright tail of the X-ray detected CCCP stellar population in Figure~\ref{NIRCMD}. The OB candidates fall into three main groups: (1) coinciding with the more reddened half of the known OB population ($A_V=2--5$~mag); (2) a narrow, vertical sequence near $J-H=0.5$ ($A_V= 6$~mag); and (3) an even more obscured population scattering to redder colors. These extinction values assume dereddening to the OB main sequence, and indeed the dereddened $J$ magnitudes for these stars would be significantly brighter (and hence younger) than the 1 Myr isochrone plotted. The X-ray-detected, "diskless" PMS population (black plus signs) does not extend upward to meet either the known or candidate OB stars, unlike the YSOs (red symbols), some of which have sufficient brightening due to IR excess emission to occupy the same area of the color-magnitude diagram as the (more reddened) OB candidates.

While it is possible that some intermediate-mass PMS stars can lose their disks in ${<}0.5$~Myr (see, e.g.\ \citealt{povich2016}) and become misidentified as highly reddened candidate OB stars, we do not expect this source of contamination to be as important as that from unassociated field stars. Indeed, most of the OB candidates that have been newly classified as cool stars are located far from the star-forming clusters (Figure~\ref{fig-spatial}), which is incompatible with the extremely young ages required for PMS contaminants.

We performed preliminary SED fitting as described by P11 for 18 of the 21 confirmed OB candidates listed in Table 3 (omitting three candidates for which we were unable to assign spectral subclasses).  With our new spectral classifications available, we estimated $T_{\rm eff}$ and fixed this parameter, allowing interstellar extinction $A_V$ (governed by the "anomalous" average extinction law with $R_V = 4$ used by P11) and bolometric luminosity $L_{\rm bol}$ to remain free parameters.  These parameters are listed in Table 3.  Our group is currently performing a more quantitative spectroscopic analysis of these stars, and improved values of $T_{\rm eff}$, $A_V$, and $L_{\rm bol}$ are forthcoming in a future paper.

The confirmed new OB stars range in $A_V$ from 1.5 to 11 mag, with a median value of 3.4 mag. The median extinction of the new OB stars is higher than 90\% of the previously cataloged OB stars in Carina measured by the same SED-fitting method (P11), which helps to explain why these stars were missed by previous visible-light surveys. Bolometric luminosities in general agree well with the reported spectral types. The most significant outlier is OBc 3, with $\log{L_{\rm bol}/L_{\sun}}=4.48$, about 0.5 dex lower than expected for a main-sequence O7 V star. Given the location of this star on the outskirts of the Carina Nebula (Fig 1), it is possible that it belongs to a background star cluster and we hence underestimated its distance and luminosity.

Most of the confirmed OB stars are distributed away from the clusters; only OBc 24 and 71 lie close to known clusters. As a measure of clustering, we used a nearest-neighbors analysis to determine the local surface number density of OB stars.  
\citet{ch85} give the local \textit{mass} surface density estimator $\mu_{j}$ as
\begin{equation}
\mu_{j} = \frac{j-1}{S(s_{j})}m
\end{equation}
where $S(s_{j}) = \pi r_{j}^{2}$ for the $j^{th}$-nearest neighbor and $m$ is the mass of each star. We opted to eliminate $m$ from the calculation in order to remove one element of uncertainty from the calculation, reducing the value to the surface number density. We also chose $j~=~6$ as this represents a good compromise between resolution and accuracy in the surface density estimator \citep{ch85}.

\begin{figure}[htp]
\epsscale{1.1}
\plotone{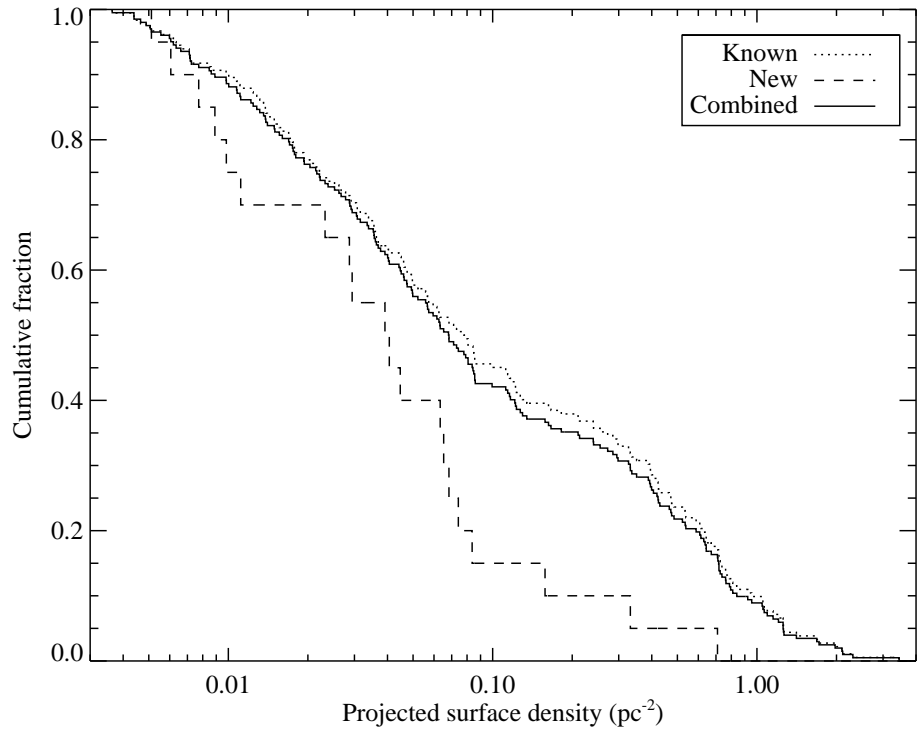}
\caption{Cumulative histograms of projected surface number densities for previously known OB stars (dotted line), newly confirmed candidate OBs (dashed line), and combined known/new OB stars (solid line). \label{fig-nn}}
\end{figure}

Figure~\ref{fig-nn} is a cumulative histogram of the projected surface number density (pc$^{-2}$) for Carina OB stars, where the conversion to parsecs assumed a 2.3 kpc distance to Carina (P11).The solid line is the combined distribution of densities for the entire sample (Tables \ref{table-obk} and \ref{table-obcob}), the dotted line represents the surface densities for ``known'' OB stars (Table~\ref{table-obk}), and the dashed line is for newly confirmed massive stars (Table~\ref{table-obcob}).

The distributions of the previously known and newly confirmed OB stars are clearly very different, with nearly 30\% of the known OB stars above 0.35 pc$^{-2}$, while there are no newly confirmed OB stars above this threshold. This simple fact should not be surprising given the low probability of detecting new OB stars within extremely well-studied clusters like Tr~14, 15, and 16.  Therefore, we were surprised to discover a new O5 V/O7 III--V star (OBc 50) on the outskirts of the obscured cluster Tr 16-SE (Sanchawala et al. 2007).  Of our confirmed new OB stars, it is the most luminous and corresponds to Tr 16-MJ 568 \citep{massey1993}. In spite of its relative proximity to Tr 16 and Eta Car, MJ 568 was unclassified before now due to its relatively high extinction. 

We also discovered a new B1 V star, OBc 24, just outside the extensively-studied Tr 14 cluster. In fact, this region is heavily confused and may represent a separate subcluster outside of Tr 14.  OBc 24 may be the star Tr 14-MJ 91 \citep{massey1993}, which has no spectral type published. It is so close to Tr 14-MJ 92, an O7.5 Vz star (SIMBAD), and CPD $-$58 2608, an ambiguous B star, that we could not have put AAT fibers on all stars simultaneously.  

It is also interesting to note that the distribution of new OB stars tends toward the main star-forming ridge, with the exception of a few stars at very low surface densities.  It remains an open question whether the outlying stars OBc 17 and 89 may have formed in relative isolation with respect to other OB stars/clusters, or if they were once cluster members that were ejected as runaways.  On the other hand, the outlier ERO 39 is unlikely to be a runaway because its bow shock points toward the center of Carina rather than away \citep{sps15}, suggesting that if this star has a high space velocity, it is actually moving in the wrong direction for a runaway.

One of the earliest-confirmed O stars in our sample (OBc 3, O6/7 V) appears particularly isolated in the boondocks of western Carina.  This star is in the middle of the region studied by \citet{kumar2014} but was overlooked by them, although they did detect two previously known, nearby O stars (which they typed O8.5 III and an O9.7 V).  As they showed, OBc 3 is not actually isolated; rather, there is a young star cluster associated with it that was not identified by \citet{feigelson2011} because it was on the edge of their field of view.  It is possible that the cluster hosting OBc 3 is actually an unassociated, background cluster, which would be consistent with the anomalously low $\log{L_{\rm bol}/L_{\odot}}= 4.48$ (Table 3) for this star at our assumed 2.3 kpc Carina distance. The main-sequence luminosity of $\log{L_{\rm bol}/L_{\odot}}\approx 5$ for an O7 V star would place the star at a significantly greater distance than even the 2.9 kpc assumed by \citet{kumar2014}.

The confirmation of OBc 59 was intriguing to us since it appears to be associated with an X-ray+YSO star cluster deeply embedded in a small IR dark cloud near the base of the ``Great Pillar'' (one of the few such high-absorption features in Carina; \citealt{feigelson2011}).  This is an exciting confirmation of a new massive cluster lurking in the depths of the Great Pillar.

With the confirmation of the three late O or early B stars OBc 75, 77, and 89/ERO 38, we have discovered a new site of massive star formation at the southeast extremity of the Carina Nebula, out beyond the rest of the action in the South Pillars.  The star OBc 77 does appear to lie within a cluster revealed by the X-ray point-source population \citep{feigelson2011}.  There is also a concentration of YSOs and 24 \micron\ nebulosity associated with these three stars.    

In summary, we have presented a spectroscopic search for new OB stars in the Carina Nebula.  While our confirmation rate was lower than P11 suspected, we have shown that new pockets of active star formation are deeply embedded throughout the region.  Most of the newly confirmed OB stars have higher extinction and appear relatively isolated, away from the well-studied open clusters.  In a forthcoming paper, we will present more quantitative analysis of the physical parameters of the OB population.  We also recommend future studies of the region to study the properties of binaries, the kinematics of the cluster, the runaway status of the isolated OB stars, and the stellar content of new open clusters across the region.  The Carina Nebula continues to be an exciting location to study massive star formation and dynamics.

\acknowledgments

The authors would like to thank the referee for several helpful comments that improved this manuscript.  
The authors also thank Sarah Mastell, Iraklis Kontantopoulus, and Angel Lopez-Sanchez for help with the AAT observations. M.J.A. and M.V.M. are supported by the National Science Foundation under grant AST-1109247.  R.J.H., M.J.A., and M.V.M. have also received institutional support from Lehigh University.  M.S.P. is grateful to the NSF for support via AST-1411851 and CAREER-1454333. AAT observations were awarded through a shared time agreement between the NOAO and AAO.  This research has made use of the SIMBAD database, operated at CDS, Strasbourg, France, and the WEBDA database, operated at the Department of Theoretical Physics and Astrophysics of the Masaryk University, Czech Republic.


\medskip

\bibliography{references}

\begin{thebibliography}{}
\expandafter\ifx\csname natexlab\endcsname\relax\def\natexlab#1{#1}\fi

\bibitem[{{Casertano} \& {Hut}(1985)}]{ch85}
{Casertano}, S., \& {Hut}, P. 1985, \apj, 298, 80

\bibitem[Feigelson et al.(2011)]{feigelson2011} 
Feigelson, E.~D., Getman, K.~V., Townsley, L.~K., et al.\ 2011, \apjs, 194, 9 

\bibitem[{{Gagn{\'e}} {et~al.}(2011){Gagn{\'e}}, {Fehon}, {Savoy}, {Cohen},
  {Townsley}, {Broos}, {Povich}, {Corcoran}, {Walborn}, {Remage Evans},
  {Moffat}, {Naz{\'e}}, \& {Oskinova}}]{gfs11}
{Gagn{\'e}}, M., {Fehon}, G., {Savoy}, M.~R., {et~al.} 2011, \apjs, 194, 5

\bibitem[{{Getman} {et~al.}(2011){Getman}, {Broos}, {Feigelson}, {Townsley},
  {Povich}, {Garmire}, {Montmerle}, {Yonekura}, \& {Fukui}}]{gbf11}
{Getman}, K.~V., {Broos}, P.~S., {Feigelson}, E.~D., {et~al.} 2011, \apjs, 194,
  3
  
\bibitem[Kumar et al.(2014)]{kumar2014} 
Kumar, B., Sharma, S., Manfroid, J., et al.\ 2014, \aap, 567, A109 

\bibitem[{{Levato} \& {Malaroda}(1981)}]{lm81}
{Levato}, H., \& {Malaroda}, S. 1981, \pasp, 93, 714

\bibitem[Levato et al.(1991)]{levato1991} 
Levato, H., Malaroda, S., Garcia, B., et al.\ 1991, \apss, 183, 147

\bibitem[{{Massey} {et~al.}(2001){Massey}, {DeGioia-Eastwood}, \&
  {Waterhouse}}]{mdw01}
{Massey}, P., {DeGioia-Eastwood}, K., \& {Waterhouse}, E. 2001, \aj, 121, 1050

\bibitem[{{Massey} \& {Johnson}(1993)}]{massey1993}
{Massey}, P., \& {Johnson}, J. 1993, \aj, 105, 980

\bibitem[{{Povich} {et~al.}(2011{\natexlab{a}}){Povich}, {Smith}, {Majewski},
  {Getman}, {Townsley}, {Babler}, {Broos}, {Indebetouw}, {Meade}, {Robitaille},
  {Stassun}, {Whitney}, {Yonekura}, \& {Fukui}}]{psm11}
{Povich}, M.~S., {Smith}, N., {Majewski}, S.~R., {et~al.} 2011{\natexlab{a}},
  \apjs, 194, 14

\bibitem[{{Povich} {et~al.}(2011{\natexlab{b}}){Povich}, {Townsley}, {Broos},
  {Gagn{\'e}}, {Babler}, {Indebetouw}, {Majewski}, {Meade}, {Getman},
  {Robitaille}, \& {Townsend}}]{ptb11}
{Povich}, M.~S., {Townsley}, L.~K., {Broos}, P.~S., {et~al.}
  2011{\natexlab{b}}, \apjs, 194, 6  (P11)

\bibitem[Povich et al.(2016)]{povich2016} 
Povich, M.~S., Townsley, L.~K., Robitaille, T.~P., et al.\ 2016, \apj, 825, 125

\bibitem[Sanchawala et al.(2007)]{sanchawala2007} 
Sanchawala, K., Chen, W.-P., Lee, H.-T., et al.\ 2007, \apj, 656, 462 

\bibitem[{{Sexton} {et~al.}(2015){Sexton}, {Povich}, {Smith}, {Babler},
  {Meade}, \& {Rudolph}}]{sps15}
{Sexton}, R.~O., {Povich}, M.~S., {Smith}, N., {et~al.} 2015, \mnras, 446, 1047

\bibitem[Siess et al.(2000)]{siess2000} 
Siess, L., Dufour, E., \& Forestini, M.\ 2000, \aap, 358, 593

\bibitem[{{Smith} {et~al.}(2010){Smith}, {Povich}, {Whitney}, {Churchwell},
  {Babler}, {Meade}, {Bally}, {Gehrz}, {Robitaille}, \& {Stassun}}]{spw10}
{Smith}, N., {Povich}, M.~S., {Whitney}, B.~A., {et~al.} 2010, \mnras, 406, 952

\bibitem[Sota et al.(2014)]{sota2014} 
Sota, A., Ma{\'{\i}}z Apell{\'a}niz, J., Morrell, N.~I., et al.\ 2014, \apjs, 211, 10 

\bibitem[{{Townsley} {et~al.}(2011){Townsley}, {Broos}, {Corcoran},
  {Feigelson}, {Gagn{\'e}}, {Montmerle}, {Oey}, {Smith}, {Garmire}, {Getman},
  {Povich}, {Remage Evans}, {Naz{\'e}}, {Parkin}, {Preibisch}, {Wang}, {Wolk},
  {Chu}, {Cohen}, {Gruendl}, {Hamaguchi}, {King}, {Mac Low}, {McCaughrean},
  {Moffat}, {Oskinova}, {Pittard}, {Stassun}, {ud-Doula}, {Walborn}, {Waldron},
  {Churchwell}, {Nichols}, {Owocki}, \& {Schulz}}]{tbc11}
{Townsley}, L.~K., {Broos}, P.~S., {Corcoran}, M.~F., {et~al.} 2011, \apjs,
  194, 1

\bibitem[{{Walborn} \& {Fitzpatrick}(1990)}]{wf90}
{Walborn}, N.~R., \& {Fitzpatrick}, E.~L. 1990, \pasp, 102, 379

\end{thebibliography}
\bibliographystyle{apj} 

\newpage


\clearpage

\begin{deluxetable}{lcccccll}
\tabletypesize{\scriptsize}
\tablecaption{Sample of Known Carina Massive Stars \label{table-obk}}
\tablewidth{0pt}
\tablehead{\colhead{Name} & \colhead{RA} & \colhead{DEC} & \colhead{ST$_\mathrm{4100}$} & \colhead{ST$_\mathrm{4375/4410}$} & \colhead{ST$_\mathrm{lit}$} & \colhead{Notes} \\
            \colhead{(1)} & \colhead{(2)} & \colhead{(3)} & \colhead{(4)} & \colhead{(5)} & \colhead{(6)} & \colhead{(7)}
}
\startdata
HD 92607   & 10:40:12.43 & -59:48:10.10 &   O8.5 V + O9 V   &  O9 V + O9 V &   O9 II/III   &  SB2\tablenotemark{a}, ERO 36\tablenotemark{g} \\
HD 305443   & 10:40:30.10 & -59:56:51.20 &   B0.5 V   &  B2 III  &   B2 III   &   \nodata \\
HD 92644   & 10:40:31.71 & -59:46:43.90 &   O9.5 V   &  B0 V  &   O9.5 V / B0 III   &   \\
LS 1745   & 10:40:39.25 & -60:05:36.10 &   B0.5 V   &  B0.5 V  &   B2 III   &   \nodata \\
LS 1760   & 10:41:15.30 & -59:57:43.00 &   B1 III   &  B2 III  &   B2 Ib   &   \nodata \\
LS 1763   & 10:41:20.29 & -60:06:36.40 &   B3-5 V   &  B2 V  &   B2 V   &  \nodata \\
HD 303225   & 10:41:35.48 & -59:39:45.00 &   B1.5 V   &  B1.5 V  &   B1.5 V   &   \nodata \\
HD 92852   & 10:41:54.20 & -59:06:36.50 &   B1.5 V   &  B2 V  &   B1 V   &  \nodata \\
HD 303202   & 10:41:55.83 & -59:16:16.70 &   B1 V   &  B1 V  &   B3 III   &   \nodata \\
HD 303189   & 10:41:59.21 & -59:07:49.70 &   B1 V   &  B0.5 V  &   B2 V   &   \nodata \\
HD 305452   & 10:42:02.31 & -60:08:38.60 &   B0 V   &  B2 III  &   B2 III   &   \nodata  \\
HD 92877   & 10:42:07.50 & -59:54:24.10 &   B0.5 V   &  B0.5 V  &   B2 III   &   Double/multiple star\tablenotemark{b} \\
HD 303297   & 10:42:13.34 & -59:09:46.80 &   B0 V   &  B1 III  &   B1 V   &   \nodata \\
HD 92894   & 10:42:15.83 & -59:08:14.00 &   B0 V   &  B1 III  &   B0 IV   &   \nodata \\
HD 303296   & 10:42:25.04 & -59:09:24.50 &   B0 V   &  B1 III  &   B1 Ve   &   \nodata \\
HD 92937   & 10:42:32.15 & -59:35:30.30 &   B2 V   &  B3 V  &   B2-3 II   &   \nodata \\
Coll 228-30   & 10:42:36.16 & -59:59:26.20 &   O9.5 V   &  B0.5 V  &   B1.5 V\tablenotemark{c}   &   \nodata \\
HD 305438   & 10:42:43.78 & -59:54:16.50 &   O8.5 I   &  O9 III  &   O8 Vz   &   \nodata \\
HD 305437   & 10:42:45.18 & -59:52:19.60 &   B0 V   &  B0 V  &   B0.5 V   &   \nodata \\
HD 303313   & 10:42:50.18 & -59:25:31.30 &   B1.5 V + B2 V  &  B2 V + B2 V  &   B1.5 V   &   SB2\tablenotemark{a} \\
HD 305535   & 10:42:54.64 & -59:58:19.70 &   B3-5 V   &  B2.5 V  &   B2.5 Vn   &   \nodata \\
HD 93002   & 10:43:04.15 & -59:04:59.80 &   B0.5 V   &  B1 III  &   B2 III   &   \nodata \\
HD 303316   & 10:43:11.18 & -59:44:21.00 &   O8 III   &  O7.5 V  &   O7 V((f))z   &   \nodata \\
HD 93028   & 10:43:15.35 & -60:12:04.20 &   O9 V   &  B0 V  &   O9 IV   &   SB\tablenotemark{b} \\
HD 93026   & 10:43:16.33 & -59:10:27.20 &   B1 V   &  B1 V  &   B2 III   &   \nodata \\
HD 93027   & 10:43:17.96 & -60:08:03.20 &   O9.5 V   &  B0 III  &   O9.5 IV   &   \nodata \\
HD 93056   & 10:43:27.40 & -60:05:54.80 &   B0 V + B2 V   &  O9 V + B2 V  &   B2/5   &   SB2\tablenotemark{a}  \\
HD 303312   & 10:43:30.82 & -59:29:23.80 &   O9.5 V   &  B0 V  &   O9.7 IV   &   EB\tablenotemark{b} \\
Tr 14-30   & 10:43:33.35 & -59:35:11.10 &   B0 III   &  B1 II  &   B1 Ia   &   \nodata \\
HD 305556   & 10:43:43.41 & -60:20:27.40 &   O9.7 I   &  B1 Ib  &   B0 Ib   &   \nodata \\
Tr 14-28   & 10:43:43.56 & -59:34:03.50 &   B1 V   &  B2 III  &   B2 V   &   \nodata \\

Tr 14-27   & 10:43:43.89 & -59:33:46.20 &   B1.5 V   &  B1.5 V  &   O9 V   &   \nodata \\
HD 305518   & 10:43:44.00 & -59:48:17.90 &   O9 V   &  B0 V  &   O9.7 III   &   \nodata \\
LS 1813   & 10:43:45.08 & -59:53:25.40 &   B1 V   &  B0.5 V  &   B2 Vb   &   \nodata \\
Tr 14-20   & 10:43:46.69 & -59:32:54.90 &   ON8 V   &  \nodata  &   O6 V((f))z  &   SB\tablenotemark{d} \\
HD 93097   & 10:43:46.99 & -60:05:49.20 &   B0 V   &  B0.5 IV  &   A2   &   \nodata \\
Tr 14-21   & 10:43:48.71 & -59:33:24.10 &   \nodata   &  O9 V  &   O9 V   &   SB\tablenotemark{e} \\
Tr 14-22   & 10:43:48.81 & -59:33:35.20 &   \nodata   &  B2 V  &   B2 V   &  \nodata \\
Coll 228-48   & 10:43:48.87 & -60:09:00.90 &   B2 V   &  B2 V  &   B1.5 Vb:\tablenotemark{c} &   \nodata \\
HD 305521   & 10:43:49.40 & -59:57:22.70 &   B0.5 V   &  B0 V  &   B0.5 Vn   &   \nodata \\
Tr 14-24   & 10:43:50.90 & -59:33:50.60 &   B1 V   &  \nodata  &   B1 V   &  \nodata \\
HD 93128  & 10:43:54.41 & -59:32:57.42 &   O3.5 V((f))   &  O5-6 V  &   O3.5 V((fc))z   &  SB\tablenotemark{d} \\
LS 1821   & 10:43:57.49 & -60:05:28.10 &   O9 V   &  B0 V  &   O9 V   &   \nodata \\
Tr 14-18   & 10:43:57.96 & -59:33:53.70 &   B1 V   &  B2 V  &   B1.5 V   &   \nodata \\
Tr 14-19   & 10:43:58.46 & -59:33:01.60 &   B3-5 V   &  B5 V  &   B1 V   &   \nodata \\
Coll 228-68   & 10:44:00.13 & -60:06:07.50 &   B1 V   &  B0 V  &   B1 Vn\tablenotemark{c}  &   \nodata \\
HD 93130   & 10:44:00.35 & -59:52:27.90 &   O6.5 V   &  O8 V  &   O6.5 III(f)   &   EB\tablenotemark{b} \\
LS 1822   & 10:44:00.62 & -59:25:49.20 &   B0 V   &  B1 III  &   B1.5 Ib   &   \nodata \\
Tr 16-127   & 10:44:00.93 & -59:35:45.90 &   O9 V   &  O9 V  &   O9.2 V   &   \nodata \\
Tr 14-29   & 10:44:05.11 & -59:33:41.47 &   B2 V   &  B1 V  &   B1.5 V   &   ERO 21\tablenotemark{g} \\  
Tr 16-124   & 10:44:05.83 & -59:35:11.70 &   B0 V   &  B1 V  &   B1 V   &   \nodata \\
HD 305520   & 10:44:05.83 & -59:59:41.70 &   B0 III   &  B1 Ia &   B1 Ib   &   \nodata \\
HD 305536   & 10:44:11.05 & -60:03:21.70 &   O9 V   &  B0.5 V  &   O9.5 V   &   SB\tablenotemark{b} \\
Tr 16-245   & 10:44:13.80 & -59:42:57.10 &   B0 V + B1 V:    &  \nodata  &   B0 V  & SB2\tablenotemark{a}   \\
Tr 16-246   & 10:44:14.75 & -59:42:51.80 &   \nodata   &  B1 V  &   B0.5 V   &   \nodata \\
HD 305522   & 10:44:14.97 & -60:00:05.70 &   B0 V   &  B0 V + B1 V  &   B0.5 V   &   SB2\tablenotemark{a}  \\
LS 1837   & 10:44:15.87 & -60:09:04.40 &   B1.5 V   &  B0.5 V  &   B1 V   &   \nodata \\
Tr 16-11   & 10:44:22.52 & -59:39:25.80 &   B1.5 V   &  B3 V  &   B1.5 V   &  \nodata \\
LS 1840   & 10:44:24.62 & -59:30:35.90 &   B1.5 V   &  B1 V + B2 V  &   B1 V?   &   SB2\tablenotemark{a}  \\
Tr 16-122   & 10:44:25.49 & -59:33:09.30 &   B2 V   &  B1.5 V  &   B1.5 V   &   \nodata \\
Tr 16-94   & 10:44:26.47 & -59:41:02.90 &   B1.5 V   &  B1.5 V  &   B1.5 V   &   \nodata \\
Tr 16-18   & 10:44:28.97 & -59:42:34.30 &   B2 V   &  B2 V  &   B2 V   &   \\

Tr 15-26   & 10:44:29.09 & -59:20:04.90 &   B0 V   &  B1 III  &   B1 V   &   \nodata \\
Tr 16-12   & 10:44:29.42 & -59:38:38.10 &   B0 V   &  B1 III  &   B1 V   &   \nodata \\
HD 305523   & 10:44:29.42 & -59:57:18.30 &   O9 I   &  O9 V  &   O9 II-III   &   \nodata \\
Tr 16-10   & 10:44:30.37 & -59:37:26.70 &   B0.5 V   &  O7 V + O8 V   &   B0 V   &   SB2\tablenotemark{a}  \\
Tr 16-17   & 10:44:30.49 & -59:41:40.60 &   B0.5 V   &  B1 Vp  &   B1 V  &   \nodata \\
Tr 15-23   & 10:44:30.77 & -59:21:26.00 &   B0.5 V   &  B2 III  &   B0 V   &   \nodata \\
HD 93204   & 10:44:32.34 & -59:44:31.00 &   O6 V   &  O7 V  &   O5.5 V((f))   &   \nodata \\
Tr 16-13   & 10:44:32.90 & -59:40:26.10 &   B2 V   &  B1 V  &   B1 V   &   \nodata \\
Tr 15-19   & 10:44:35.92 & -59:23:35.60 &   B1 V   &  B2 V  &   O9 V   &   \\
HD 93222   & 10:44:36.24 & -60:05:29.00 &   O8 V   &  O8 V  &   O7 V((f))z   &   \nodata \\
Tr 15-18   & 10:44:36.36 & -59:24:20.30 &   B1 Ib   &  B1 Ia  &   O9 I/II:e::  &   \nodata \\
Tr 16-21   & 10:44:36.69 & -59:47:29.60 &   O9.5 V   &  O9.5 V  &   O8.5 V + B0.5: V:  &   SB2\tablenotemark{f} \\
LS 1853   & 10:44:36.79 & -59:54:24.80 &   B2 Ia   &  B1 Ib  &   B1 Ib   &   \nodata \\
Coll 228-36   & 10:44:36.87 & -60:01:11.70 &   B2 V   &  B2 V  &   B0.5 V + B0.5 V:\tablenotemark{c}  &   SB2\tablenotemark{c} \\
Tr 16-14   & 10:44:37.19 & -59:40:01.50 &   B0 V   &  B1 V  &   B0.5 V  &   \nodata \\
HD 303311   & 10:44:37.46 & -59:32:55.40 &   O8 III((f))   &  O7 V  &   O6 V((f))z   &   Double/multiple star\tablenotemark{b} \\
Tr 15-21   & 10:44:37.66 & -59:23:07.30 &   B0 V   &  B1 V:  &  B0 III   &   \nodata  \\
Tr 16-20   & 10:44:38.65 & -59:48:14.10 &   B2 V   &  B2 V  &   B1 V   &   \nodata \\
Tr 16-16   & 10:44:40.31 & -59:41:49.00 &   B0.5 V   &  B0 V  &   B1 V   &   \nodata \\
Tr 15-14   & 10:44:40.57 & -59:22:28.40 &   B2 V   &  \nodata  &   B2.5 IV-V   &   \nodata \\
Tr 16-15   & 10:44:40.98 & -59:40:10.40 &   B0 V   &  B1 III  &   B0 V   &   \nodata \\
Tr 16-100   & 10:44:41.78 & -59:46:56.70 &   O6 V   &  O8 V  &   O6 V((f))   &   \nodata \\
Tr 15-15   & 10:44:42.34 & -59:23:03.90 &   \nodata   &  B1 V  &   B0.5 IV-V   &   \nodata \\
Tr 15-9   & 10:44:42.35 & -59:22:03.00 &   B2 V   &  B2 V  &   B1 V   &   \nodata \\
Tr 15-2   & 10:44:43.73 & -59:21:16.30 &   B0 V   &  \nodata  &   O9.5 III:   &   \nodata \\
HD 93249 A   & 10:44:43.80 & -59:21:24.20 &   \nodata   &  B0 V  &   O9 III   &   \nodata \\
HD 305524   & 10:44:45.24 & -59:54:41.50 &   O5/6 V((f))   &  O7-8 V  &   O6.5 Vn((f))z   &   \nodata \\
Tr 16-104   & 10:44:47.31 & -59:43:53.20 &   O8.5 V   &  O7 V + O8 V  &   O7.5 V(n)z + B0 V(n)  &   SB2\tablenotemark{f} \\
HD 305534   & 10:44:47.51 & -59:57:58.90 &   B3: V + ?   &  B0 V + B0 V  &   B0.5 V: + B1 V\tablenotemark{c}   &   SB2\tablenotemark{c} \\
Tr 16-29   & 10:44:53.76 & -59:37:48.30 &   B2 V   &  B2 III  &   B2 V   &   \nodata \\
Tr 16-5   & 10:44:54.08 & -59:41:29.40 &   B1 V   &  B3 V  &   B1 V   &   \nodata \\

LS 1865   & 10:44:54.71 & -59:56:01.90 &   O8.5 V((f))   &  O9 V  &   O8.5 V   &   \nodata \\
Tr 16-31   & 10:44:56.29 & -59:33:03.50 &   B0 V   &  B2 V  &   B0.5 V   &   \nodata \\
LS 1866   & 10:44:57.33 & -60:00:46.80 &   B0.5 V   &  B1 III  &   B2 V   &   \nodata \\
Tr 16-26   & 10:44:59.90 & -59:43:14.80 &   B2 V   &  \nodata  &   B1.5 V  &   \nodata \\
Tr 16-25   & 10:45:00.24 & -59:43:34.40 &   \nodata   &  B2 V  &   B2 V   &   \nodata \\
Tr 16-4   & 10:45:02.16 & -59:42:01.00 &   B0.5 V   &  B1 V  &   B1 V  &   \nodata \\
Tr 16-23   & 10:45:05.79 & -59:45:19.60 &   O8.5 V   &  B0 V  &   O7.5 V(n)z   &  \nodata \\
Tr 16-9   & 10:45:05.84 & -59:43:07.70 &   O9.5 V   &  B2 I + B0-1 V   &   O9.7 IV   &   SB2\tablenotemark{a}  \\
Tr 16-24   & 10:45:05.88 & -59:44:18.90 &   B2 V   &  B2 V  &   B2 V   &   \nodata \\
HD 303308   & 10:45:05.85 & -59:40:06.40 &   O4 V((f))   &  O5-6 V  &   O4.5 V((fc))   &   \nodata \\
Tr 16-3   & 10:45:06.73 & -59:41:56.50 &   O9.5 V   &  B0.5 V  &   O9.5 V   &   \nodata \\
Tr 16-1   & 10:45:08.23 & -59:40:49.50 &   O9.5 V   &  B0 V + B0 V   &   O9.5 V(n) + B0.5 V(n)   &  SB2\tablenotemark{d}  \\
Tr 16-22   & 10:45:08.22 & -59:46:07.00 &   O9 V   &  O8 V  &   O8.5 Vp   &   \nodata \\
Tr 16-74   & 10:45:09.74 & -59:42:57.20 &   B1 V   &  B1 V  &   B1 V   &   \nodata \\
Tr 16-2   & 10:45:11.20 & -59:41:11.30 &   B1 V   &  B1 V &   B1 V   &   \nodata \\
HD 93343   & 10:45:12.22 & -59:45:00.40 &   O8.5 V   &  \nodata  &   O8 Vz   &   \nodata \\
Tr 16-110   & 10:45:12.87 & -59:44:19.20 &   O8.5 V &  B0 V + B0 V  &   O8 V + O8 V   &  SB2\tablenotemark{f}  \\
HD 305533   & 10:45:13.37 & -59:57:53.80 &   B1 V   &  B0 V  &   B0.5 Vnn   &   \nodata \\
Tr 16-112   & 10:45:16.52 & -59:43:37.00 &   O6/7 V   &  O7 V  &   O6 V((fc))   &   SB2\tablenotemark{d} \\
HD 305528   & 10:45:16.73 & -59:54:45.80 &   B2 V  &  B2 V  &   B2 V   &   \nodata \\
HD 93342   & 10:45:17.57 & -59:23:37.50 &   B1 I   &  B1 Ia  &   O9 III   &   \nodata \\
Tr 16-55   & 10:45:19.43 & -59:39:37.40 &   B2 V   &  B2 V  &  B1.5 V  &   \nodata \\
Tr 16-115   & 10:45:20.57 & -59:42:51.30 &   OC9 V   &  B0 V  &  O9 V   &   \nodata \\
Tr 16-28   & 10:45:22.14 & -59:37:38.50 &   B1 V   &  B2 III  &   B2 V  &   \nodata \\
HD 305532   & 10:45:34.06 & -59:57:26.70 &   O8.5 III   &  O8 V  &   O6.5 V((f))z   &   \nodata \\
FO 15   & 10:45:36.32 & -59:48:23.20 &   O5 V:   &  O5 V:  &  O5 Vz + B0: V  &   EB\tablenotemark{b} \\
HD 305538   & 10:45:46.45 & -60:05:13.30 &   O9.5 V   &  B0.5 V  &   B0 V   &   Binary/multiple star\tablenotemark{b} \\
Coll 228-81   & 10:45:53.45 & -60:05:37.10 &   B1 V   &  B0.5 V  &  B0.5 Vb\tablenotemark{c}  &   \nodata \\
HD 305525   & 10:46:05.70 & -59:50:49.30 &   O4/5 V((f))   &  O8 V  &   O5.5 V(n)((f))z   &   \nodata \\
HD 93501   & 10:46:22.03 & -60:01:18.90 &   B1 V   &  B0 V  &   B2 II/III   &   \nodata \\
LS 1892   & 10:46:22.46 & -59:53:20.40 &   O4/5 V((f))   &  O6 V  &   O5.5 V(n)((f))z   &   \nodata \\

LS 1893   & 10:46:25.41 & -60:08:43.70 &   O9.5 V   &  B1 III  &   B0 V   &   \nodata \\
HD 305539   & 10:46:33.07 & -60:04:12.60 &   O8.5 V   &  O9 V  &   O8 Vz   &   \nodata \\
HD 303304   & 10:46:35.69 & -59:37:00.60 &   O6 V   &  O7-8 V  &   O7 V   &  \nodata \\
HD 93576   & 10:46:53.84 & -60:04:41.90 &   O9.5 V   &  B0 V  &   O9.5 IV(n)   &   SB\tablenotemark{b} \\
HD 93620   & 10:47:09.20 & -59:47:29.80 &   B2 V   &  B2 III  &   B2 II/III   &    \\
HD 93632   & 10:47:12.49 & -60:05:49.80 &   O5 V((f))   &  O6.5 V  &   O5 Ifvar   &   \nodata \\
Coll 228-89   & 10:47:13.18 & -60:13:33.80 &   B1.5 V   &  B2 V  &   B2 V\tablenotemark{c}   &   \nodata \\
HD 303402   & 10:47:19.20 & -59:27:33.50 &   B1 V   &  B1 V  &   B1 V   &   \nodata \\
LS 1914   & 10:47:22.08 & -60:05:57.70 &   O9.5: V   &  B2 V  &   B2 Ve   &   \nodata \\
HD 93683   & 10:47:38.89 & -60:37:04.45   &   O9 V   &  O9 V + B0 V &   B0/1 Vne &  SB2\tablenotemark{a}, ERO 37\tablenotemark{g}  \\  
HD 93723   & 10:48:01.67 & -59:39:03.20 &   B2 V   &  B2 V  &   B3 III   &   \nodata \\
HD 305619   & 10:48:15.52 & -60:15:56.70 &   O9 I   &  O9 III  &   O9.7 II   &   \nodata \\
HD 303413   & 10:48:58.95 & -59:41:09.10 &   B0 III   &  B1 III  &   B1 Ib   &   \nodata \\
HD 305602   & 10:49:08.91 & -59:53:28.10 &   B1 V   &  B2 V  &   B2 V   &   \nodata \\
HD 93911   & 10:49:13.64 & -60:11:03.30 &   B3 Ia   &  B2 II  &   B2/3 Iab/b   &   \nodata \\
HD 305599   & 10:49:24.98 & -59:49:43.70 &   O9.5 V   &  B0 V  &   B0 Ib   &   \nodata \\
HD 305606   & 10:49:25.84 & -60:01:37.10 &   B1.5 V   &  B2 V  &   B2 V   &   \nodata \\
\enddata
\tablecomments{Spectral types are from Simbad unless otherwise noted.  EB -- eclipsing binary; SB -- spectroscopic binary; SB2 -- double lined spectroscopic binary}
\tablenotetext{a}{This work}
\tablenotetext{b}{Simbad}
\tablenotetext{c}{\citealt{lm81}}
\tablenotetext{d}{\citealt{levato1991}}
\tablenotetext{e}{WEBDA}
\tablenotetext{f}{\citealt{sota2014}}
\tablenotetext{g}{\citealt{sps15}}
\end{deluxetable}


\clearpage
\begin{deluxetable}{lccccclccc}
\tabletypesize{\scriptsize}
\tablecaption{Spectroscopically Confirmed Massive Star Candidates \label{table-obcob}}
\tablewidth{0pt}
\tablehead{\colhead{OBc} & \colhead{RA} & \colhead{DEC} & 
           \colhead{ST$_\mathrm{4100}$} & \colhead{ST$_\mathrm{4375/4410}$} & \colhead{Notes} &  $T_{\rm eff}$  (K)  &  $\log{L_{\rm bol}/L_{\sun}}$  &  $A_V$ (mag)   \\
            \colhead{(1)} & \colhead{(2)} & \colhead{(3)} & \colhead{(4)} & \colhead{(5)} & \colhead{(6)}  &  \colhead{(7)}  &  \colhead{(8)}  &  \colhead{(9)} 
}
\startdata
3 & 10:40:23.46 & -59:50:39.00 &    O6 V:    &    O7 V    &  \nodata  &  36.5   &    4.48    &   3.38  \\
17 & 10:42:44.85 & -60:05:02.00 &    B0 V    &    B1 V    &    \nodata   &  28.0   &    4.62   &    3.36  \\
23 & 10:43:40.56 & -60:10:03.40 &    B3 V    &    B2 V    &    \nodata   &  19.0   &    3.73   &    2.47   \\
24 & 10:43:41.21 & -59:35:53.30 &    B1 V    &    B1  V    &  \nodata   &  26.0    &   4.60    &   4.03  \\
38 & 10:44:30.26 & -59:26:12.90 &    O9.5 V    &    B0 V    &   \nodata  &  30.5   &    4.69   &    3.43   \\
39 & 10:44:30.89 & -59:14:46.00 &    ND    &    O8 V:    &    \nodata   &  35.0   &    4.83    &   5.39  \\
40 & 10:44:52.43 & -60:16:10.60 &    B0 V: + B2 V    &    B2 V    &    SB2  &  28.0   &    4.27    &   2.71  \\
47    & 10:45:11.18 & -59:42:33.80 &    ND    &    B5 V:    &    \nodata  &  15.0    &   3.54   &    11.03  \\
49 & 10:45:20.42 & -59:17:06.10 &    B0.5 V    &    B1 V    &    \nodata  &  27.0   &    4.72   &    3.35   \\
50 & 10:45:22.29 & -59:50:47.00 &    O5 V:     &    O7 III-V    &    \nodata  &  37.0   &    5.46   &    4.90  \\
53 & 10:45:31.95 & -60:00:29.10 &    ND    &    O9 V    &    \nodata  &  33.0   &    4.73   &    6.06   \\
55 & 10:45:36.45 & -59:44:10.70 &    ND    &    O9 V    &    \nodata  &  33.0   &    5.13    &   7.04   \\
57 & 10:45:38.70 & -60:04:26.50 &    B1.5 V    &    B2 V    &    \nodata  &  22.5  &     3.90   &    2.08  \\
59 & 10:45:41.93 & -60:16:52.10 &    ND    &    B:    &    \nodata &  \nodata  &  \nodata  &  \nodata  \\
60 & 10:45:45.36 & -59:58:53.00 &    B1 III    &    B1 III    &    \nodata  &  26.0   &    4.12   &    4.90  \\
61 & 10:45:46.64 & -59:48:40.20 &    ND    &    B:   &    \nodata &  \nodata  &  \nodata  &  \nodata  \\
68 & 10:46:19.06 & -59:57:54.30 &    B0 III    &    B1 III   &    \nodata  &  28.0   &    4.39   &    2.96  \\
71 & 10:46:52.25 & -60:06:03.30 &    B0.5 V    &    B0.5 IV    &    \nodata &  28.0   &    4.58   &    3.53   \\
75 & 10:47:35.26 & -60:29:23.40 &    O9.5 III    &   B1 III    &    \nodata  &  31.0   &    4.45   &    2.59  \\
77 & 10:47:37.35 & -60:26:26.01 &    ND    &    B:    &    \nodata &  \nodata  &  \nodata  &  \nodata  \\
89 & 10:49:14.26 & -60:15:56.80 &    B2 V    &    B2 III    &    ERO 38 &  21.0   &    3.90   &    1.51  \\
ERO 18   & 10:43:20.29 & -60:13:01.35 &    B8 V    &   B8 V &  \nodata  &  \nodata  &  \nodata  &  \nodata    \\  
ERO 39  &  10:48:46.53 &  -60:35:39.9  &  B0 V   &  B0.5 III &  \nodata  &  \nodata  &  \nodata   &  \nodata   \\   
\enddata
\tablecomments{SB2 -- double lined spectroscopic binary}
\end{deluxetable}

\begin{deluxetable}{lccccl}
\tabletypesize{\scriptsize}
\tablecaption{Spectroscopically Rejected or Undetected Massive Star Candidates \label{table-obcfgk}}
\tablewidth{0pt}
\tablehead{\colhead{OBc} & \colhead{RA} & \colhead{DEC} & 
           \colhead{ST$_\mathrm{4100}$} & \colhead{ST$_\mathrm{4375/4410}$} \\
            \colhead{(1)} & \colhead{(2)} & \colhead{(3)} & \colhead{(4)} & \colhead{(5)}
}
\startdata
1 & 10:39:09.94 & -59:47:14.50 &   ND   &   K0-K5 III   \\
2 & 10:40:14.67 & -59:56:54.40 &   ND   &   G0-K0 I   \\
4  & 10:40:44.15 & -59:49:34.80 &  ND   &   ND   \\
5 & 10:40:59.29 & -59:27:24.90 &   F   &   \nodata   \\
6 & 10:41:09.05 & -59:44:00.50 &   ND   &   G0-K0 I   \\
7 & 10:41:18.94 & -59:35:44.50 &   ND   &   G0-K0 I   \\
8 & 10:41:25.96 & -59:23:19.60 &   ND   &   ND   \\
9 & 10:41:51.03 & -60:02:39.10 &   ND   &   G0-K0 I   \\
10 & 10:41:54.91 & -59:41:23.60 &   ND   &   G0-K0 I   \\
11 & 10:42:04.67 & -59:54:57.20 &   ND   &   G0-K0 I   \\
12 & 10:42:05.01 & -59:53:17.40 &   ND   &   G0-K0 I   \\
13 & 10:42:06.81 & -59:37:01.70 &   ND   &   G0-K0 I   \\
14 & 10:42:14.48 & -59:59:45.40 &   ND   &   G0-K0 I   \\
15 & 10:42:31.48 & -59:24:44.70 &   GK   &   \nodata   \\
16 & 10:42:39.15 & -59:28:16.00 &   A5   &   \nodata   \\
18 & 10:42:46.53 & -60:12:07.00 &   F   &   \nodata   \\
19  & 10:43:06.95 & -59:19:15.00 &   ND   &   ND   \\
20 & 10:43:25.19 & -59:12:14.70 &  ND   &   G0-K0 I   \\
21 & 10:43:26.46 & -59:37:24.90 &   A3-5   &   \nodata   \\
22 & 10:43:40.38 & -60:17:04.00 &   ND   &  G5-K5 I   \\
25 & 10:43:44.72 & -60:27:05.50 &   ND   &   K I   \\
26 & 10:43:45.96 & -59:29:33.90 &   ND   &   G0-K0 I   \\
27 & 10:43:47.75 & -59:30:56.80 &   ND   &   ND   \\
28  & 10:43:49.96 & -59:36:55.60 &   ND   &   ND   \\
29  & 10:44:00.19 & -60:14:14.00 &   ND   &   ND   \\
30 & 10:44:01.63 & -59:03:27.40 &   ND   &   G0-K0 I   \\
31 & 10:44:01.80 & -59:24:34.20 &   F   &   \nodata   \\
32 & 10:44:02.75 & -59:39:46.00 &   ND   &   G5-K5 I   \\
33 & 10:44:04.07 & -60:00:17.10 &   ND   &   G5-K5 I   \\
34 & 10:44:11.16 & -59:52:42.60 &   ND   &   G5-K5 I   \\
35 & 10:44:18.03 & -60:02:51.70 &   ND   &   ND   \\
36  & 10:44:26.06 & -59:27:37.04 &   ND   &   ND   \\
37 & 10:44:27.81 & -59:45:21.40 &   ND   &   G0-K0 I   \\
41 & 10:44:57.51 & -59:54:29.50 &   ND   &   G0-K0 I   \\
42  & 10:45:00.90 & -59:47:18.70 &  ND   &   ND   \\
43 & 10:45:03.61 & -59:18:15.30 &   ND   &   G0-K0 I   \\
44 & 10:45:04.33 & -59:54:35.20 &   ND   
&   G0-K0 I   \\
45  & 10:45:07.24 & -60:14:26.00 &   ND   &   ND   \\
46 & 10:45:07.50 & -59:33:44.90 &   ND   &   G0-K0 I   \\
48  & 10:45:17.21 & -59:47:01.60 &   ND   &   ND   \\
52  & 10:45:30.22 & -59:48:21.00 &   ND   &   ND   \\
54  & 10:45:33.12 & -60:25:04.20 &   ND   &   F I  \\
56 & 10:45:36.75 & -59:47:02.20 &   ND   &   ND   \\
58  & 10:45:40.66 & -59:53:45.00 &   ND   &   ND   \\
62  & 10:45:47.07 & -60:05:18.50 &   ND   &   ND   \\
63 & 10:45:49.10 & -59:25:42.30 &   F   &   \nodata   \\
64 & 10:45:53.14 & -59:44:40.50 &   ND   &   ND   \\
65 & 10:45:54.99 & -60:12:12.10 &   ND   &   G5-K5 I   \\
66  & 10:45:57.88 & -60:07:07.30 &   ND   &   ND   \\
67 & 10:46:15.19 & -59:32:17.60 &   F   &   \nodata   \\
69 & 10:46:38.67 & -60:16:13.30 &   F5-G0   &   \nodata   \\
70   & 10:46:38.75 & -59:36:07.80 &   ND   &   \nodata   \\
72 & 10:46:58.97 & -60:17:21.40 &   ND   &   G0-K0 I   \\
73 & 10:47:02.99 & -59:50:19.20 &   ND   &   G5-K5 I   \\
74 & 10:47:14.56 & -60:17:37.70 &   ND   &   G5-K5 I   \\
76 & 10:47:35.79 & -59:29:53.70 &   ND   &   G0-K0 I   \\
78  & 10:47:38.07 & -60:07:24.80 &   ND   &   K:   \\
79  & 10:47:46.41 & -60:24:36.00 &   ND   &   ND   \\
80   & 10:47:54.55 & -60:08:00.00 &   ND   &   \nodata   \\
81 & 10:47:58.74 & -59:45:17.30 &   ND   &   G5-K5 I  \\
82 & 10:48:11.96 & -59:56:06.60 &   ND   &   G5-K5 I   \\
83 & 10:48:26.24 & -59:59:12.60 &   ND   &   K0-K5 V   \\
84 & 10:48:26.62 & -59:30:20.10 &   F   &   \nodata   \\
85 & 10:48:27.47 & -59:29:06.00 &   F   &   \nodata   \\
87 & 10:48:38.92 & -59:30:02.00 &   A5   &   \nodata   \\
88 & 10:48:58.62 & -59:50:57.40 &   F   &   \nodata   \\
90 & 10:49:19.78 & -59:43:37.70 &   A3   &   \nodata   \\
91 & 10:49:22.78 & -59:48:06.00 &   F0-3   &   \nodata   \\
92   & 10:49:23.73 & -60:09:56.90 &   ND   &   \nodata   \\
93 & 10:49:28.17 & -59:43:55.90 &   ND   &   G0-K0 I    \\
94 & 10:42:20.83 & -59:09:08.60 &   F   &   \nodata   \\
\enddata
\tablecomments{ND -- non-detection}  
\end{deluxetable}

\end{document}